\documentclass[preprint,10pt]{elsarticle}

\usepackage{natbib}
\setcitestyle{authoryear,open={(},close={)}} 

\usepackage{amssymb}
\usepackage{amsmath}

\journal{Journal of Environmental Management}

\begin{document}

\begin{frontmatter}



\title{Assessing the potential impact of environmental land management schemes on emergent infectious disease risks}


\author[1]{Christopher~J.~Banks}
\author[2]{Katherine~Simpson}
\author[2]{Nick~Hanley}
\author[1,3,4]{Rowland~R.~Kao}

\affiliation[1]{organization={Roslin Institute, University of Edinburgh},
            country={United Kingdom}}
\affiliation[2]{organization={School of Biodiversity, One Health and Veterinary Medicine, University of Glasgow},
             country={United Kingdom}}
\affiliation[3]{organization={School of Physics and Astronomy, University of Edinburgh},
  country={United Kingdom}}
\affiliation[4]{Corresponding author: rowland.kao@ed.ac.uk}

\begin{abstract}
  Financial incentives encourage the plantation of new woodland to increase habitat, biodiversity, carbon sequestration, as a contribution to meeting climate change and biodiversity conservation targets. Whilst these are largely positive effects, it is worth considering that this expansion of woodland can lead to increased presence of wildlife species in proximity to agricultural holdings that may pose an enhanced risk of disease transmission between wildlife and livestock. Wildlife and the provision of a reservoir for infectious disease is particularly important in the transmission dynamics of bovine tuberculosis, the case studied here.
  
  In this paper we develop an economic model for predicting changes in land use resulting from subsidies for woodland planting. We use this to assess the consequent impact on wild deer populations in the newly created woodland areas, and thus the emergent infectious disease risk arising from the proximity of new and existing wild deer populations and existing cattle holdings.
  
  We consider an area in the South-West of Scotland, having existing woodland, deer populations, and extensive and diverse cattle farm holdings. In this area we find that, with a varying level of subsidy and plausible new woodland creation scenarios, the contact risk between areas of wild deer and cattle increases between 26\% and 35\% over the risk present with a zero subsidy.
  
  This provides a foundation for extending to larger regions and for examining potential risk mitigation strategies, for example the targeting of subsidy in low disease risk areas, or provisioning for buffer zones between woodland and agricultural holdings.
\end{abstract}




  



\end{frontmatter}



\section*{Introduction}
Economic incentives are often used to encourage private landholders to undertake costly actions that benefit biodiversity \citep{Kangas2019}. Examples include Agri-Environment-Climate policies under Pillar 2 of the Common Agricultural Policy, and the recently launched Environmental Land Management Scheme in England. The Scottish Government's Forestry Grant Scheme \citep{scottishgovernment2019} encourages landholders to change their land management to encourage the plantation of new woodland on land that is currently used for arable and/or livestock farming. This has a positive benefit for the creation of new woodland habitat, increasing some measures of biodiversity, and improving carbon sequestration (depending on location), as well as providing economic benefits for the land owner, such as the ability to sell carbon credits attached to new woodland planting. However, one potential social cost of this expanded woodland cover is that the resultant increase in both the density and variety of wildlife will change the patterns of their contact with livestock and humans, potentially increasing disease risks to both populations. In particular, woodland planting subsidies provided to agricultural landowners potentially bring woodland species closer to agricultural holdings.

Emerging infectious disease risks present both an economic burden to livestock keepers and are a potential threat to human and animal health and welfare. One source of disease concern in many countries around the world is the impact of increasing populations of wild deer, with recent examples that under appropriate conditions, deer species have proven capable hosts for the emergence of Chronic Wasting Disease \citep{Otero2021}, contributing to the maintenance of bovine tuberculosis (bTB) \citep{Salvador2019,Ward2009}, and spreading of SARS-CoV-2 \citep{Caserta2023}. As deer numbers and density are likely to increase in northern countries with the increase in forests due to climate change policies and biodiversity target-setting, their potential role to contribute to disease threats may be considerable. 

As the threat of infectious diseases may in turn influence the management choices that land managers will make in the future and the determination of the socially optimal pattern of land use, understanding how land management changes affect those contact patterns has implications for both disease and land management. A critical step in understanding these changes is quantifying the change in disease risks and tracking them over time. This paper provides such an analysis for the specific case of bTB in South-West Scotland. We consider the increased risk of bTB maintenance by wild deer that come into contact with agricultural cattle holdings. Badgers are the better know wildlife host for bTB \citep{Gallagher2000} and in Scotland few deer have been recorded with bTB infections. However, in the USA white tailed deer act as a reservoir for bTB \citep{OBrien2023a} and red deer (which are one of the most common deer species in Scotland) act as a maintenance host for bTB in Austria and the Alpine regions \citep{Fink2015}. A recent sample of red deer in Exmoor (in England's bTB High Risk Area) found 28.3\% to be positive for bTB \citep{Collard2023}.

In this analysis, we concentrate on the problem as it relates to cattle-deer interactions, as this is a concern for bovine tuberculosis maintenance due to increased potential for transmission between wild deer and cattle. A critical issue is that, while increases in woodland planting may increase the density of deer, it is the geographical distribution of both new and existing woodland, which determines the extent to which deer and cattle are likely to increase contact as well. This is the focus of our analysis. We build upon prior work \citep{Simpson2021,Simpson2021a} evaluating land cover changes under agri-environment scheme incentives, to characterise potential land use change under various levels of incentives. We then use previous estimates of deer density dependence on land cover usage to assess the likely distribution of deer in relation to existing and new woodland plantation. We then consider these new deer populations in relation to the location of cattle, and how contacts between deer and cattle are likely to change with new plantation. Thus, we identify what is the likely impact of subsidies for new woodland on infection risks, using the adjacency of deer and cattle populations as an indicator of inter-species disease transmission risk.

\section*{Methods and data}
To implement the model, data is needed on livestock density and current distribution, deer densities and distributions, and economic returns from alternative land uses.

\subsection*{Data}
\subsubsection*{Land use}
Our analysis is based on a case study landscape in the South-West of Scotland in an area of mixed land use and therefore a plausible location for examining the impact of changing land use with farmers being incentivised to convert land currently being farmed to woodland. Original land use data was taken from the UK Centre for Ecology and Hydrology (UKCEH) Land Cover maps \citep{Marston2022} and Land Cover Plus: Crops \citep{UKCEH2022}. From these datasets we extracted the set of land parcels in the study area along with their most recently designated land use type. Land parcels in the UKCEH Land Parcel Spatial Framework vary in size, but have a minimum area of 0.5 hectares. They represent discrete real-world units of land dominated by a single land cover type \citep{Smith2007}. The dataset used is from year 2022.

The study area chosen was a 30km by 20km area of Dumfries and Galloway, as an area with existing woodland, wild deer populations, and both extensive and intensive cattle farming. In the model, the initial land use type is assigned from the land cover map with any arable areas broken down to crop types from the land cover plus crops map. Table~\ref{tab:land-use-props} details the land cover types in the study area and the proportion of area covered by each.

\begin{table}
  \centering
  \begin{tabular}{|l|r|r|r|}
    \hline
    Land cover type & Land parcels & Area (km$^2$) & \% by area\\
    \hline
    Improved grassland & 6512 & 248.5 & 46.3\\
    Coniferous woodland & 1571 & 95.5 & 17.8\\
    Semi-natural grassland & 703 & 69.8 & 13.0\\
    Mountain/heath/bog & 1301 & 49.4 & 9.2\\
    Deciduous woodland & 2692 & 47.3 & 8.8\\
    Freshwater & 81 & 11.5 & 2.1\\
    Built-up & 566 & 5.1 & 0.9\\
    Arable (Grass) & 72 & 4.0 & 0.7\\
    Arable (Spring barley) & 34 & 2.2 & 0.4\\
    Arable (Other crops) & 40 & 1.4 & 0.3\\
    Arable (Winter barley) & 18 & 1.2 & 0.2\\
    Arable (Winter wheat) & 11 & 0.8 & 0.1\\
    Arable (Maize) & 4 & 0.3 & 0.1\\
    Arable (Spring field beans) & 2 & 0.1 & 0.0\\
    Arable (Oilseed rape) & 1 & 0.1 & 0.0\\
    Arable (Potatoes) & 1 & 0.0 & 0.0\\
    \hline
  \end{tabular}
  \caption{Land use in the study area, by land cover area.}
  \label{tab:land-use-props}
\end{table}

\subsubsection*{Cattle}
Information on the location of cattle holdings and distribution of cattle were extracted from the Cattle Tracing System (CTS) of the British Cattle Movement Service (BCMS) \citep{UKGovernment2022} and the National Agricultural Census (agCensus) \citep{EDINA2023}.

All cattle holdings within the study area were extracted from the CTS location database and their most recent cattle populations extracted by examining their registered cattle movements in the CTS movement database. Cattle at each holding were then distributed to the holding location's nearest ``improved grassland'' land parcels in proportion with the agCensus cattle density at each land parcel. Figure~\ref{fig:animal-dist} (left) shows the animal populations distributed to land parcels in the model.

There were around 36,000 cattle in the study area at the latest available date in the CTS location database, October 2022.

\subsubsection*{Deer}
Estimated deer populations in the study area were extracted from raster data generated using a refinement of the process described in Croft et al. \citep{Croft2017} and aggregated to the land parcel areas using the rasterio \citep{Rasterio2023} and geopandas \citep{GeoPandas2023} Python packages. Red and roe deer were considered to be the only species with significant presence in the study area. Figure~\ref{fig:animal-dist}(left) shows the animal populations distributed to land parcels in the model.

\subsubsection*{Current agricultural returns and woodland planting subsidies}
For each of the current agricultural land use types in the study area, a value for gross margin per hectare was taken from the Farm Management Handbook \citep{farmmanagementhandbook202223pdfa}. Gross margins show the profit associated with each crop/livestock land use option, defined as income from sales net of variable costs (such as fertiliser, seeds and pesticides). Gross margin values in the handbook are quoted with sensitivity to various input and output types (e.g.\ grain price and yield). For this model we took the central estimates to give a typical margin for the land type. For cattle we assume all herds are moderate input dairy cattle as a typical gross margin value for cattle grazing land. Table~\ref{tab:margin-subsidy} lists the gross margin values used in the model and the initial fixed value of the woodland subsidy.

\begin{table}
  \centering
  \begin{tabular}{|l|r|}
    \hline
    Land type/subsidy & Margin/subsidy \\
    \hline
    Oilseed rape & £1,340 \\
    Potatoes & £1,393 \\
    Spring barley & £773 \\
    Spring field beans & £835 \\
    Spring oats & £422 \\
    Winter barley & £1,079 \\
    Winter wheat & £1,315 \\
    Grass (without livestock) & £0 \\
    Grass (with livestock) & £4,865\\
    Maize & £$\infty$ \\
    Other crops & £$\infty$ \\
    \hline
    Woodland plantation subsidy & £1,104 \\
    \hline
  \end{tabular}
  \caption{Gross margins and subsidy for different land types, in GBP per hectare per annum. Values are central estimates from the Farm Management Handbook (FMH) \citep{farmmanagementhandbook202223pdfa}. Unimproved grassland without livestock has unknown value and we consider this to have value low enough that it will change use. FMH does not list margins for maize or other crop types and so we consider them to have value high enough they will not change use.}
  \label{tab:margin-subsidy}
\end{table}

Current forestry grant subsidies were taken from the Scottish Government Forestry Grant Scheme \citep{scottishgovernment2019}. Only the annualised subsidy level for plantation of new broadleaf woodland was considered. This was chosen both to consider a fairly central subsidy value and because deer tend to populate deciduous woodland over coniferous.

\subsection*{Model}
\subsubsection*{Land use change}
In order to model the potential for land use change due to woodland subsidy provision, the above data are first aggregated into a single dataset. For the set of land parcels in the study area we record the current land use, cattle density, and deer density. Land use types are aggregated into one of: woodland, pasture, arable, or other. Pasture is assigned a cattle density and woodland is assigned a deer density. Pasture and arable land are assigned current gross margin values, according to crop type or the presence of cattle. Figure~\ref{fig:model-landtype} (left) shows the initial land types assigned in the model.

An agent-based model, based on Simpson et al. \citep{Simpson2021a}, underlies the land use allocation process. Land use change at the parcel level is predicted based upon whether the level of the economic incentive offered for woodland planting is greater than the current agricultural gross margin of the land parcel (the landholder's opportunity cost). For land parcels that are currently pasture or arable, if the offered subsidy is greater than landholder's opportunity cost then the expectation is the landholder will sign up to the scheme and plant new woodland on that land parcel.  We model this profit maximisation process as occurring instantaneously. Note that if for any parcel the opportunity cost (current agricultural return) exceeds the planting subsidy, then the prediction is that no woodland planting will occur, and so current land use persists in that parcel.

To assess the sensitivity of land use change to subsidy, we model land manager choices across a range of subsidy levels for woodland planting from zero to double the original value, in eleven increments. From these results, we define three broader subsidy categories for comparison, low (0.5$\times$), mid (1.0$\times$), and high (1.5$\times$) multipliers relative to the baseline subsidy level (where mid is the original fixed subsidy level). These categories capture the range of plausible responses without requiring all individual steps to be analysed separately. Note that we do not include timber sales or carbon credit values as additional revenue streams for patches that switch to woodland: the only financial benefit to the land  manager of new woodland is assumed to be equal to the value of the planting subsidy.

\subsubsection*{Deer distribution}
\begin{figure}
  \centering
  \includegraphics[width=0.5\textwidth]{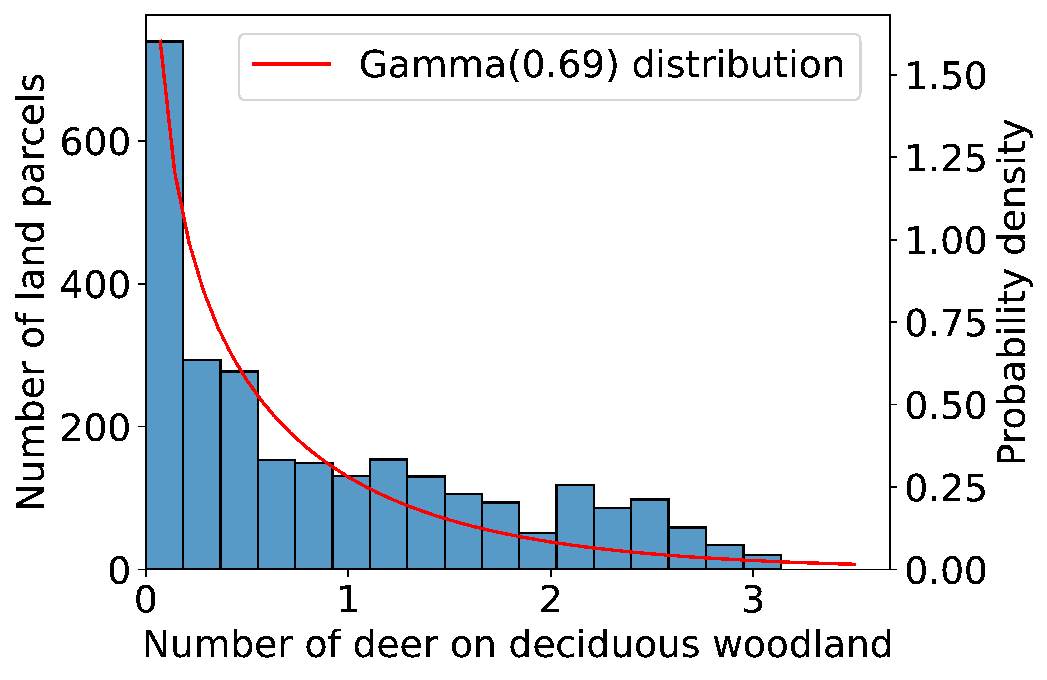}
  \caption{Distribution of the existing deer population per deciduous woodland land parcel fit to a Gamma distribution with shape parameter 0.69}
  \label{fig:gamma-dist}
\end{figure}

For land parcels with pre-existing woodland, deer populations remain the same after the land use model is applied. We then distribute new deer populations to any land parcels with newly-created woodland. New deer populations are distributed in the model according to an approximation of the pre-existing distribution of deer in deciduous woodland land parcels. Figure~\ref{fig:gamma-dist} shows the distribution of the existing deer population in deciduous woodland areas compared to a Gamma distribution with shape parameter 0.69. Deer population in new woodland land parcels is distributed by drawing from that distribution.

\subsubsection*{Animal adjacency networks}
We assess the risk model for disease transmission between wild (deer) and agricultural (cattle) species, and within species, by considering changes in the contact patterns amongst and between land parcels holding these two species, under the assumption that this network of contacts is important for establishing the risk of spread. We thus construct a two-layer network. In both networks the vertices are land parcels in the study area. In the first network (all-species network) edges exist between all spatially adjacent land parcels that contain either cattle or deer with a population estimate greater than 0.5 animals per km$^2$. In the second network (inter-species network) edges exist only between pairs of spatially adjacent land parcels where one contains cattle and the other contains a deer population estimate greater than 0.5 animals per km$^2$. The all-species network represents a mapping of all land parcels that contain animals that can potentially interact and affect disease transmission; whereas the inter-species network represents a mapping of the land parcels where there is an interface between species affecting only inter-species disease transmission.

\section*{Results}
\subsection*{Fixed subsidy level}
\begin{figure}
  \centering
  \includegraphics[width=0.49\textwidth]{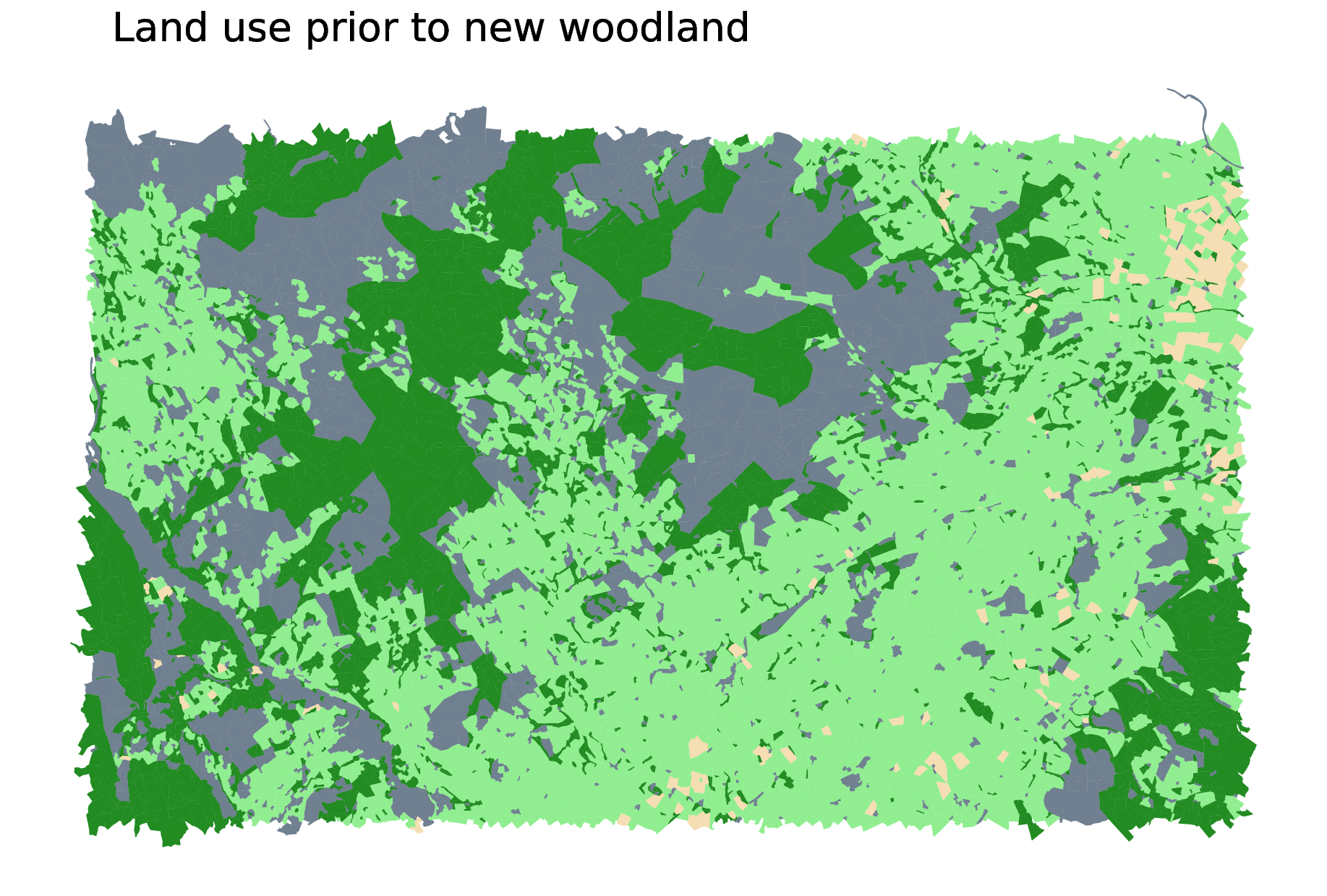}
  \includegraphics[width=0.49\textwidth]{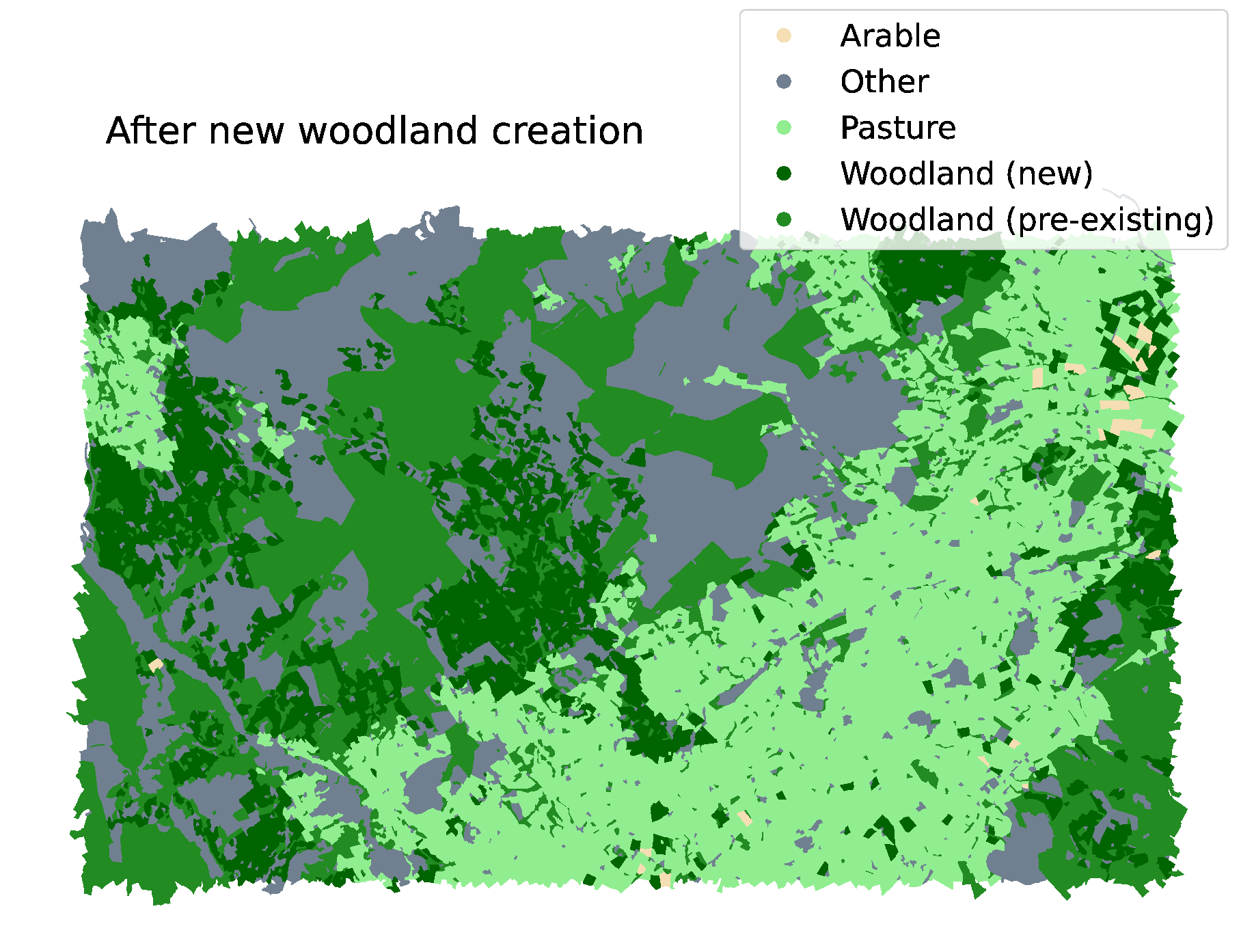}
  \caption{Land parcels prior to (left) and with new woodland created (right) in the model with a fixed subsidy level. Darkest green areas show newly created woodland. Maps show the area from British National Grid 260000--295000 E, 565000--585000 N, an area of 35 $\times$ 20 km,  oriented to Grid North, in the Dumfries and Galloway region of Scotland.}
  \label{fig:model-landtype}
\end{figure}

\begin{figure}
  \centering
  \includegraphics[width=\textwidth]{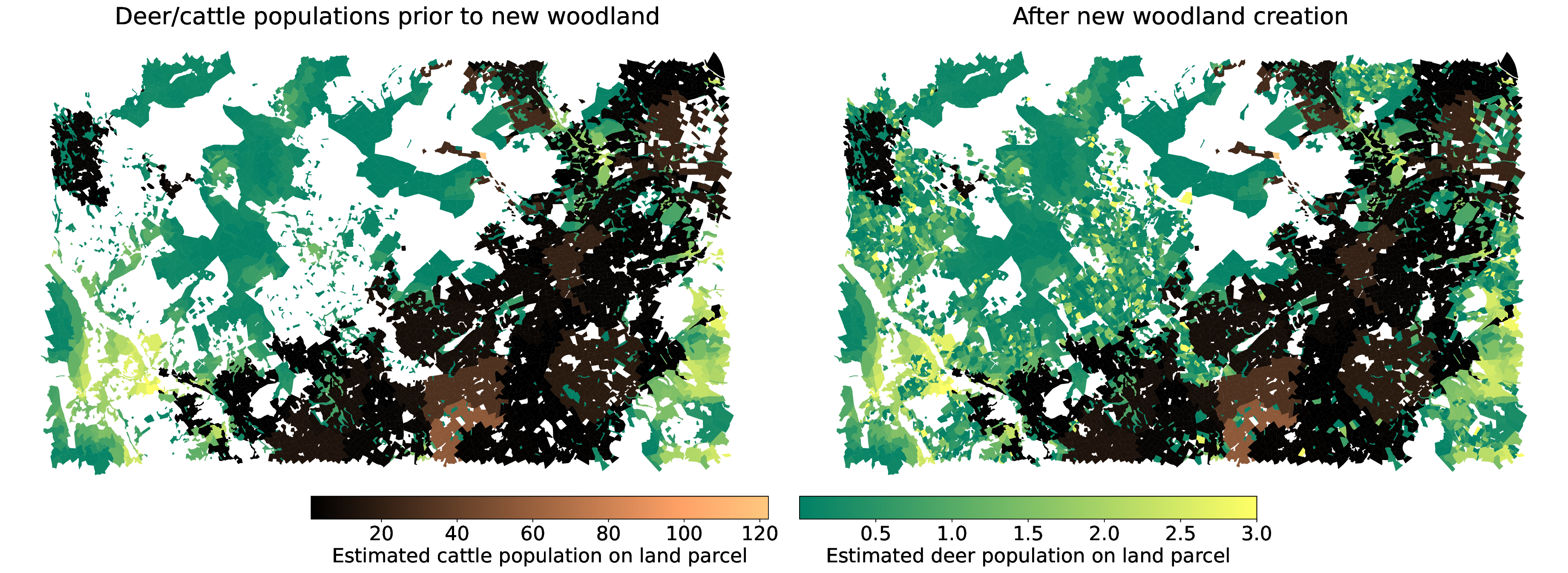}
  \caption{Estimated distribution of cattle and deer over the study area, before (left) and after (right) new woodland plantation. Colour scales denote the number of animals in each land parcel. Maps show the area from British National Grid 260000--295000 E, 565000--585000 N, an area of 35 $\times$ 20 km,  oriented to Grid North, in the Dumfries and Galloway region of Scotland.}
  \label{fig:animal-dist}
\end{figure}

Figure~\ref{fig:model-landtype} (right) details the areas of new woodland created in the model using the initial subsidy level (Table~\ref{tab:margin-subsidy}). At this level of subsidy there are 2,444 new land parcels or $\sim$80~km$^2$ of broadleaf woodland created. This results in an increase in the deer population from 2,937 to 4,607---an increase of $\sim$57\%. The cattle population of the study area is 36,538 and remains static,  as the woodland planting subsidy is not sufficient to incentivize farmers to convert profitable cattle grazing land into woodland. Figure~\ref{fig:animal-dist} (right) shows the new distribution of deer in relation to cattle after the creation of new woodland in the model.

\subsubsection*{All-species network}
After constructing the network of adjacent land parcels the result is a network with 13,609 nodes. Prior to the creation of new woodland there are 13,453 edges between land parcels with adjacent animal populations. After the creation of new woodland there are 15,378 edges between adjacent animal populations, an increase of $\sim$14.3\%.

\subsubsection*{Inter-species network}
The inter-species network only has edges between land parcels where deer and cattle are directly adjacent. Prior to the creation of new woodland there are 1,295 edges between land parcels where deer and cattle are adjacent. After the creation of new woodland there are 1,710 edges between deer and cattle, an increase of 32\%.

\begin{figure}
  \centering
  \includegraphics[width=0.45\textwidth]{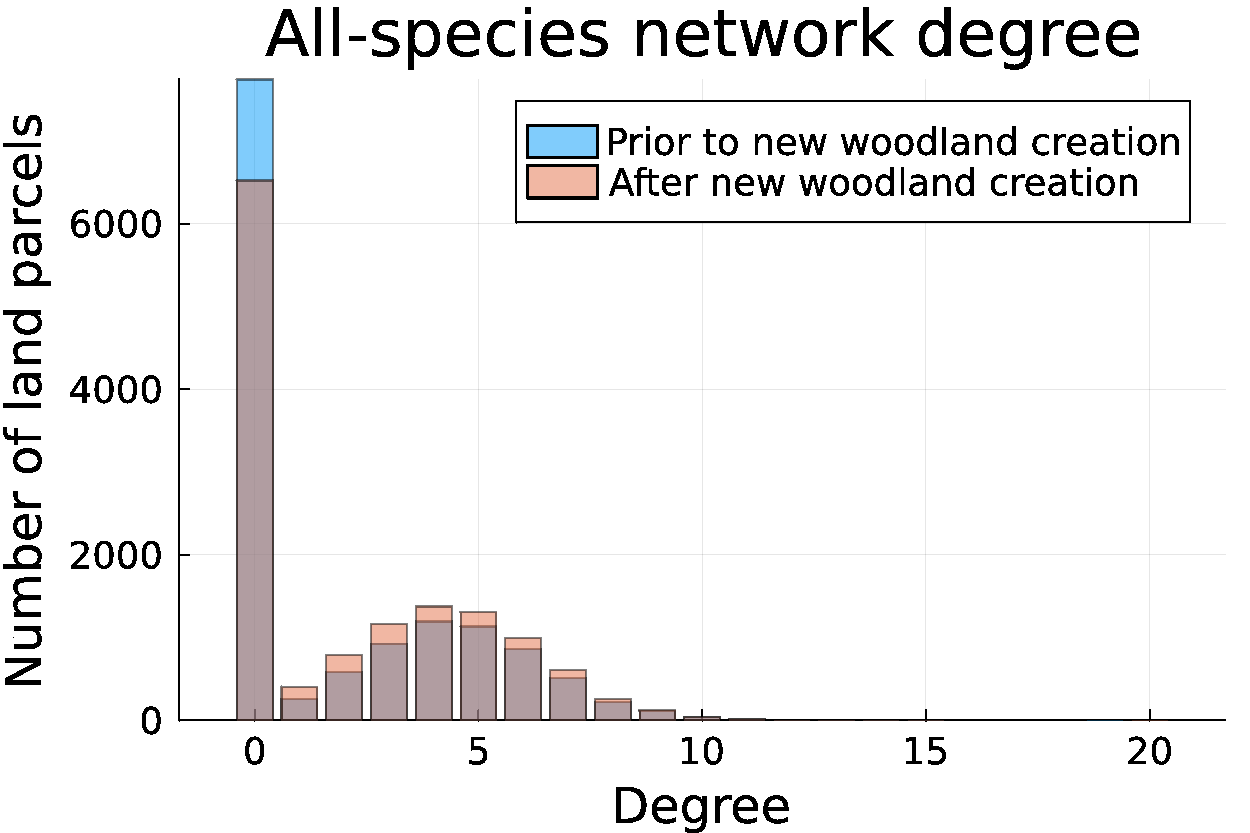}
  \hspace{1em}
  \includegraphics[width=0.45\textwidth]{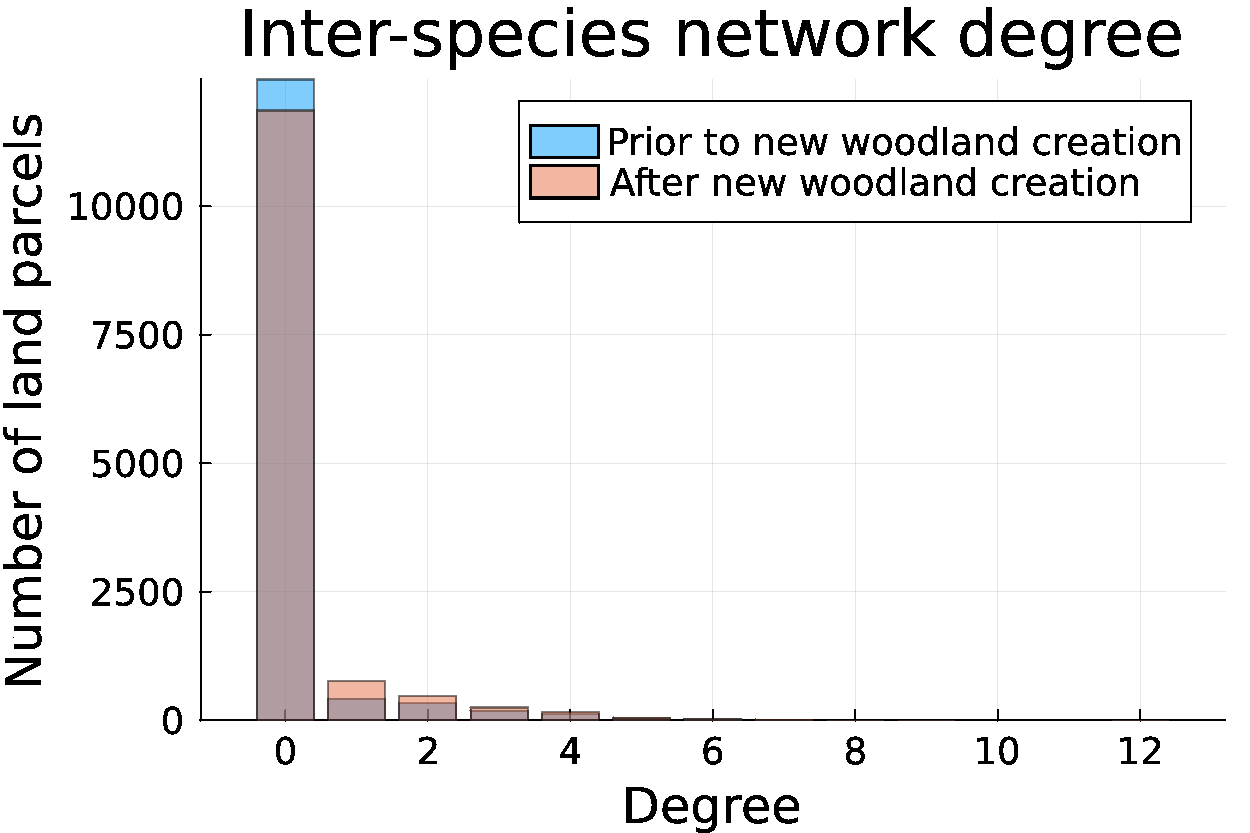}
  \caption{Adjacency degree distributions for the all-species network (left) and the inter-species network (right), prior to and post creation of new woodland. Degree is a measure of the connectivity of land parcels in each network.}
  \label{fig:degree-dists}
\end{figure}

Figure~\ref{fig:degree-dists} shows the adjacency degree distribution changes for each network. Degree is the measure of how many other land parcels a land parcel is connected to. In the inter-species network this is a measure of how many land parcels containing a species are connected to a land parcel containing a different species. In both cases we see a decrease in unconnected (degree=0) parcels and shift towards greater connectivity. In the inter-species network this indicates a greater risk of contact between species.

\subsection*{Effect of varying the subsidy level}
\begin{figure}
  \centering
  \includegraphics[width=0.45\textwidth]{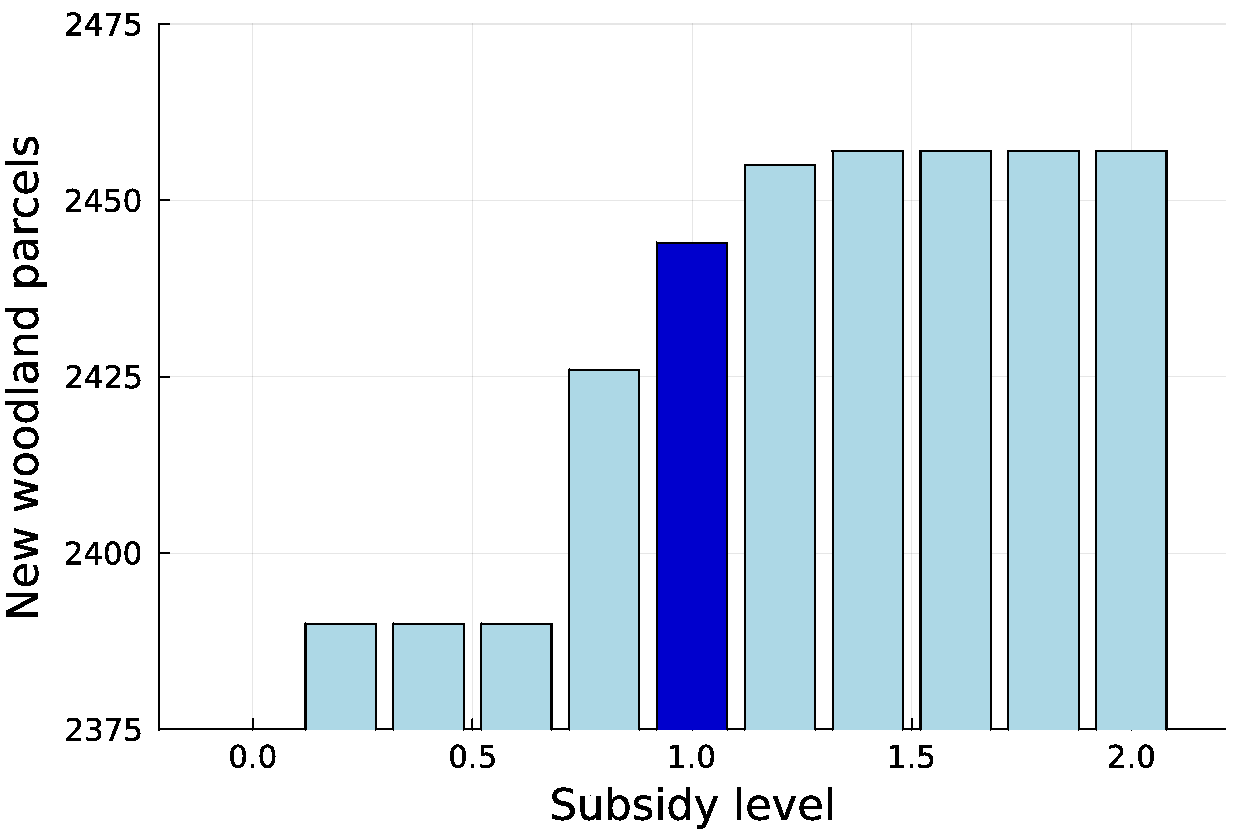}
  \hspace{1em}
  \includegraphics[width=0.45\textwidth]{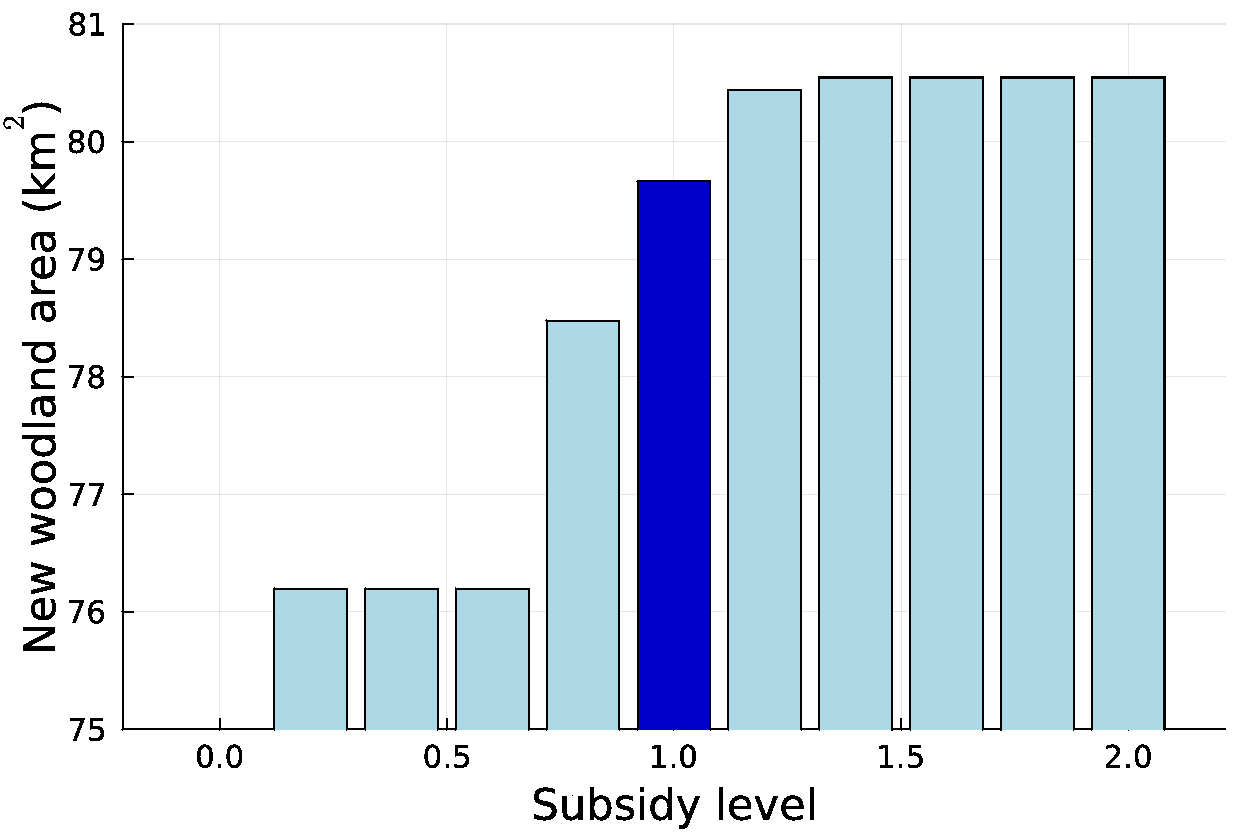}
  \caption{Sensitivity of new woodland creation to subsidy level scaling, in terms of land parcels (left) and area (right). Subsidy level 1.0 is the original fixed subsidy level.}
  \label{fig:sublevel}
\end{figure}

Varying the level of subsidy in the model from zero to double the baseline value allowed us to examine how sensitive new woodland creation is to the subsidy level. Figure~\ref{fig:sublevel} shows the sensitivity of new woodland parcels and area to a range of subsidy levels. There is a threshold region in which woodland creation responds strongly to changes in subsidy level.

For subsequent analyses, we therefore selected three subsidy levels: low, mid, and high. The low and high values correspond to the minimum and maximum levels of woodland creation observed across the full subsidy range, while the mid value corresponds to the predefined baseline subsidy. These levels correspond to 0.5, 1.0, and 1.5 times the fixed subsidy value, respectively.

Table~\ref{tab:result-lmh} details key metrics across the range of subsidy levels. The area of woodland created varies by around 3\% between the low and high subsidy scenarios, likewise the increase in deer population. However, the creation of new woodland overall (the difference between zero and high subsidy) can be seen to allow up to a 57\% increase in deer population. This results in the number of new connections between land parcels with cattle and land parcels with deer increase by up to 35\%. This is a similar effect at varying spatial scales, considering deer parcels with cattle parcels up to two or three parcels away show equivalent increases in connectivity.

\begin{table}
  \centering
  \small{
    \begin{tabular}{|l|r|r|r|r|}
      \hline
      Subsidy level & Zero & Low & Mid & High \\
      \hline
      Woodland parcels (\% increase) & 4,263 & 6,653 (56.1) & 6,707 (57.3) & 6,720 (57.6)\\
      Woodland area /km$^2$ & 142.9 & 219.1 (53.3) & 222.5 (55.8) & 223.4 (56.4)\\
      Estimated deer population & 2,937 & 4,549 (54.9) & 4,594 (56.4) & 4,616 (57.2)\\
      Deer parcels directly adjacent to cattle & 1,295 & 1,627 (25.6) & 1,710 (32.0) & 1,750 (35.1)\\
      Deer parcels up to 2 parcels away from cattle & 5,409 & 6,835 (26.4) & 7,111 (31.5) & 7,308 (35.1)\\
      Deer parcels up to 3 parcels away from cattle & 13,441 & 17,134 (27.5) & 17,827 (32.6) & 18,249 (35.8)\\
      \hline
    \end{tabular}
  }
  \caption{Key results of the model. High and Low subsidy levels correspond to the subsidy levels that create the maximum and minimum new woodland respectively. Values in parentheses are the percentage increase over the zero-subsidy scenario.}
  \label{tab:result-lmh}
\end{table}

\section*{Discussion}
In this paper we have quantified the effect of an economic incentive for new woodland planting on farmland, in terms of the enhanced risks of disease transmission between cattle and wild deer populations. We study potential changes in the between-land-parcel connections amongst cattle and deer. Our model accounts for potential land use change by adopting an agent-based model of land management decision-making in relation to the profitability of current agricultural land management and the subsidies offered for the transformation of currently-farmed patches into new broadleaved woodland. We then model a plausible distribution of wild deer in new woodland areas and their relation to existing cattle holdings. The adjacency relationship between deer and cattle populations provides us with an estimate of disease transmission risk.

\subsection*{Key assumptions}
This study is intended to be a comparative static analysis of the effect of woodland subsidy provision on the risk of inter-species disease transmission, thus we make a number of key assumptions. Apart from the change in land use from low-productivity land types to new woodland, all other land use remains static. The model output represents a state of market and biodiversity equilibrium and does not represent a temporal change over any particular timescale. Therefore, we do not consider any increase in livestock numbers or in the area of land used for livestock beyond that resulting from woodland conversion, nor do we consider any other change in land type. This affords us the ability to assess change in woodland and biodiversity in the absence of potential confounders.

Likewise, in the model, we consider woodland to be homogeneous under the assumption that we are planting woodland to maximise potential woodland biodiversity. We are not concerned with, and thus do not model, the plantation of other woodland types for other purposes. In reality the woodland subsidy would be used to plant a range of woodland types, including mono-culture forest planted for the purpose of forestry exploitation. Moreover, increases in woodland biodiversity might come at the expense of decreases in, for example, species richness on grassland.

Whilst these other types of woodland would also likely increase woodland biodiversity depending on the precise indicator selected \citep{Nthambi2024}, we are assuming the woodland planted provides the habitat desired by our deer species of interest. This also leads to the assumption that deer species take up residence in this new woodland. We essentially assume that all new woodland will be colonised by deer species to some extent. This assumption, we believe, is justifiable in the limited geographic context of this study (see below for further discussion of the geographic context). The area modelled has an existing population of deer species, and all new woodland is roughly adjacent to these populated areas, Therefore, it is reasonable to assume that new areas of idealised habitat will be colonised to some extent. This of course represents an upper bound on the new extent of deer population. In a wider study, with more diverse starting points in terms of population and habitat, due consideration would need to be made when estimating the new extent of colonisation.

We assume the basic geographic unit of the model is the land parcel. Land parcels are defined to be contiguous units of land use type, and whilst these often naturally coincide with areas of land that would be taken as a unit by the landowner, there is no reason to assume that the land may not be subdivided and changed within the land parcel unit. However, we take the land parcel as a well-defined basic unit and can propose no principled way of defining subdivisions.

Finally, we assume that an increase in adjacency between land parcels containing deer and cattle is a proxy for the increase in disease transmission risk. In the context of bTB and the role of wildlife as a reservoir and a vector for infection, we see this as a reasonable assumption. Prior analyses of the genetic distance between \emph{M.\ bovis} samples taken from cattle and wildlife \citep{Crispell2019a, Rossi2022, Crispell2020, Salvador2019} show that spatial distance between sample individuals is strongly correlated with the genetic distance between bacterial samples taken from them. As close genetic distance implies a higher likelihood of transmission links this supports the importance of spatial distance in determining infection risk.

Future extension of this work should also consider the population of each species within a land parcel and a suitable epidemiological model for infection pressure between the two species. However in this case,  in the absence of a more sophisticated epidemiological model, the connectivity between populated parcels provides a good proxy for the risk of infection pressure.

\subsection*{Extent of study}
The geographic area of this study is relatively small, covering an area of 600km$^2$, but it is an area covered by a large number of cattle farms of various sizes. It is also the case that in our economic model for land use change, we only consider average values for land use profitability at the crop/livestock type level, and we do not consider any non-economic factors in land manager decision making, such as land managers' preferences for increasing biodiversity on their land. Land managers are assumed to only care about the relative profits of alternative, mutually-exclusive land uses, and to make profitable land use changes instantaneously when the economic incentive for planting is introduced. Therefore, the results can only be considered to be an indicator of possible disease risk increase in a moderately-sized region of Scotland. However, the region has a large number of cattle farms and potentially diverse ownership and land management, changes in land use are likely to occur at a scale substantially above the size of one farm. Therefore, there is a sufficiently large population of farms (and therefore diversity of management) for our model to provide useful insights into likely responses to a woodland planting incentive.

As an exemplar for a Scottish lowland cattle farming area, and thus relevant to similar areas of cattle farming density and woodland distribution, we believe this study area serves well as a guide to the potential increase in emergent infectious disease risk from new woodland creation. More importantly, we believe that the modelling approach used here could be applied in a wide range of empirical settings where land use change results in likely changes in livestock and/or wildlife disease transmission. In general, by evaluating the adjacency between land parcels with animal species, we can assess whether fragmentation levels promote increased wildlife density, increased contact between wildlife and agriculture, and whether it has any impact on emergent infectious disease risk. Economic incentives change each of these variables in a spatially interconnected and hard to predict manner, and our model addresses these complexities.

\subsection*{Discussion of results}
Using a simple uniform planting subsidy we find that this incentive produces an increase of approximately 80 km$^2$ of new woodland in the study area. We then predict that the newly created woodlands could support a 57\% growth of the deer population. Although this absolute population increase is not the primary focus of our analysis, our interest lies chiefly in how the spatial distribution of deer relative to cattle shapes opportunities for inter connection, or proximity between deer and cattle, the magnitude is broadly consistent with expectations for a large expansion of high-quality deer habitat. Recent modelling at the UK scale similarly predicts substantial responses to woodland creation targets \citep{Hunter2025}. Roe deer and fallow deer were estimated to gain around 30\% in suitable habitat area under planting scenarios, with corresponding increases in population size (e.g. roe deer +31.3\%; fallow deer +30.1\%). These values are not directly comparable to our model outputs, but they nonetheless suggest that sizeable population growth in response to expanded woodland cover is plausible. Thus, while our estimate of a 57\% local population increase may be liberal, it is within the range of changes expected when substantial new woodland is created.

When comparing the connectivity between cattle-populated parcels and deer-populated parcels we find that this represents a 32\% increase in the number of connected land parcels and so a proportional increase in the risk of contact between these species. The caveat to this headline figure, aside from any effect of the  aforementioned modelling assumptions, is that it represents nothing more than the increase in potential for contact between species. Further investigation will be needed to asses in more detail the real epidemiological impact of this contact.

In the area of study we find that the contact risk increases between 26\% and 35\% depending on the level of subsidy provided, reflecting the corresponding expansion of woodland habitat. Notably, even with the lowest subsidy threshold that promotes woodland creation produces a significant level of increased risk. Nonetheless, the degree of risk is sensitive to subsidy level, indicating that simply adjusting subsidy level or incorporating spatial targeting could help mitigate unintended epidemiological consequences.

There is only a small increase in the total area where deer may be present and so it is unlikely that the increase in woodland area, would on its own increase the risk of deer as a wildlife reservoir. However, the substantial increase in contact between cattle and deer could dramatically increase the risk from an existing reservoir, or in the case of TB in particular, create local persistence due to circulation across the deer-cattle system. If local persistence were to result, over a sufficiently large area, this could challenge Scotland's Officially TB free status, which would be of great concern to government and industry.

\subsection*{Implications for land management}

Although varying the level of planting subsidy influences the degree of deer-cattle contact risk, the overall effect is modest, except in the case where no subsidy is provided at all. Given that the ecological and biodiversity benefits of incentivising new woodland creation are well documented \citep{Nthambi2024, Dobson2025}, the challenge is therefore to minimise disease-related risks while retaining these wider gains. Several strategies could be considered. Differentiated economic incentives could be used to encourage or discourage woodland creation in particular locations or land types (e.g.~\citep{Armsworth2012}), thereby avoiding plantation in areas immediately adjacent to agricultural land where increased deer presence could heighten transmission risks. Alternatively, spatial design principles, such as buffer zones between new woodland parcels and livestock areas may help to separate susceptible cattle from expanding deer populations. Our intention is to use the present model, and extensions of it, to explore these options further and to evaluate the potential of targeted land management policies to balance woodland creation goals with the need to limit wildlife-livestock disease risks.

\subsection*{Future work}
Future work in this area could concentrate on expanding the model to cover wider and more heterogeneous areas of land, as well as incorporating non-financial factors that influence land-use decision-making. Most importantly, however, consideration should be given to balance the need for new woodland creation (in relation to habitat formation, biodiversity, and contributions to net-zero targets) against emergent infectious disease risk. A significant addition to this will be to better represent the likely effect of this change in risk on the actual epidemiological landscape. The model should include a simulation or calculation of the actual likely disease burden and how this alters with changes in subsidy provision or management.

We have seen that, at least in this study area, the level of subsidy does not have a very large effect, as much as the binary provision of subsidy or not. This model provides a foundation for investigating other strategies for management or for  mitigating the effect of subsidy provision on disease risk. 


\section*{Acknowledgements}
We thank Theo Pepler, University of Glasgow, for provision of deer data and assistance with its processing.

\section*{Funding}
This work was supported by the Roslin Institute ISP2(theme 3), BBSRC (BBS/E/D/20002174).

\section*{License}
For the purpose of open access, the author has applied a Creative Commons Attribution (CC~BY) license to any Author Accepted Manuscript version arising from this submission.




\end{document}